\documentclass[5p]{elsarticle}
\usepackage{lineno,hyperref}
\usepackage{color}
\usepackage{longtable}
\usepackage{cleveref}

\usepackage{pdfpages}
\usepackage{subfigure}

\usepackage[inline]{enumitem}

\modulolinenumbers[5]
\usepackage{booktabs}

\journal{arXiv}

\begin{document}
\begin{frontmatter}

\title{Importance of van der Waals interactions
       in hydrogen adsorption on a silicon-carbide nanotube
       revisited with vdW-DFT and quantum Monte Carlo}

\author{Genki I. Prayogo\(^{1}\)}
\address{School of Information Science, JAIST}

\author{Hyeondeok Shin}
\address{Computational Science Division,
    Argonne National Laboratory, Argonne, Illinois 60439, USA}

\author{Anouar Benali}
\address{Computational Science Division,
    Argonne National Laboratory, Argonne, Illinois 60439, USA}

\author{Ryo Maezono}
\address{School of Information Science, JAIST}

\author{Kenta Hongo}
\address{Research Center for Advanced Computing Infrastructure, JAIST}

\begin{abstract}
  DFT is a valuable tool for calculating adsorption energies
  toward designing materials for hydrogen storage.
  However, dispersion forces being absent from the theory,
  it remains unclear how the consideration of van der Waals (vdW)
  interactions affects such calculations.
  For the first time, we applied diffusion Monte Carlo (DMC)
  to evaluate the adsorption characteristics
  of a hydrogen molecule on a (5,5) armchair silicon-carbide nanotube (H$_2$-SiCNT).
  Within the framework of density functional theory (DFT),
  we also benchmarked various exchange-correlation functionals,
  including those recently developed
  for treating dispersion or vdW interactions. 
  We found that the vdW-corrected DFT methods agree well with DMC,
  whereas the local (semilocal) functional significantly over (under)-binds.
  Furthermore, we fully optimized the H$_2$-SiCNT geometry
  within the DFT framework
  and investigated the correlation between
  structure and charge density.
  The vdW contribution to adsorption was found to be non-negligible at approximately
  1 kcal/mol per hydrogen molecule,
  which amounts to 9--29 \% of the ideal adsorption energy
  required for hydrogen storage applications.
\end{abstract}
\end{frontmatter}

\section{Introduction}

Hydrogen energy is a promising
energy resource for reducing greenhouse gas emissions~\cite{2002DIN,2009EBE,2015GRO}.
To realize the industrial use of hydrogen energy,
particularly in the transportation sector,
one of the most important developmental challenges is addressing  
the related storage issues---safety and capacity~\cite{2009EBE}.
Several materials-based strategies to store hydrogen
have been proposed,
which involve choices such as the form of the stored hydrogen
(physical vs.\ chemical storage) and the structure of the storage material
(e.g.,\ nanostructures and metal hydrides).
To achieve adsorption-based room-temperature storage,
the ideal interaction energy between the stored hydrogen and storage material
has been estimated as approximately 3.5--11.5 kcal/mol~\cite{2006JHI,2006BHA}.
At the moment, this technology is limited to low-temperature storage
because the interaction is too weak to resist being overpowered
by thermal energy at the desired higher temperatures.
Computational materials design would be immensely helpful in further
exploring appropriate storage materials from the massive materials space,
provided the hydrogen adsorption energies on candidate materials can
be accurately predicted at reasonable computational costs.

\vspace{2mm}
Silicon-carbide nanotubes (SiCNTs) are a typical nanostucture studied for the above 
purpose~\cite{2006MPO,2015DEV,2018TAB,2011WAN}
owing to their enhanced molecular interactions compared with
that of the structurally related (and more common) carbon nanotubes.
This feature has been linked to the SiCNT polarized surface
originating from the Si--C bonds~\cite{2003MAV, 2007ALA, 2007BAU, 2014BAR}. 
Despite this, studies have suggested that in the pristine state,
the adsorption energy of hydrogen on SiCNTs still lies below the required values,
ranging between 0.7 and 1.98 kcal/mol.
This has motivated further investigations into doping schemes with dopants
such as alkali and transition metals~\cite{2018TAB,2012BAN}
or with vacancies~\cite{2015DEV,2007BAI}.
These studies, however, did not incorporate van der Waals (vdW) corrections
into the conventional DFT scheme, 
and thus they may have underestimated the true adsorption potential.
From another viewpoint, the basis sets adopted for DFT calculations
also matter, since Gaussian and numerical basis sets 
can give rise to overestimated interaction energies
owing to the basis set superposition error~\cite{2011SHE}.
Thus, the true adsorption potential of SiCNTs remains unclear,
even when the results seem plausible and are consistent
with the experimentally observed higher hydrogen uptake
than that of CNTs~\cite{2014BAR2,2016BAR}.

\vspace{2mm}
In this study, we demonstrate the importance of incorporating
vdW interactions for a quantitative description of
H$_2$ adsorption on SiCNTs and related systems.
To reproduce the vdW interactions accurately,
we applied diffusion Monte Carlo (DMC) to evaluate
the adsorption energy, which can serve as an accurate reference value.
DMC has been proved to be highly accurate for
various noncovalent systems~\cite{2008KOR,2010HON,2012WAT,2013DUB,2014DUB,2013HON,
                                  2015HON,2016HON,2016DUB,2017HON,2020OQM}
and is comparable to the best correlated methods
while offering a better computational cost
and minimal systematic error~\cite{2016DUB}.
We then benchmarked several available vdW-DFT approaches
by comparing them with DMC as well as conventional exchange-correlation (XC) functionals
in terms of binding curves and charge densities
to further investigate H$_2$-SiCNT-related systems.
Finally, we investigated the fully optimized H$_2$-SiCNT geometries
for our benchmark set of XC functionals and
discussed their relationship to the charge densities at a fixed geometry,
along with the implications of their difference in terms of the DMC charge densities.

\section{Computational details}

\subsection{Structural modeling}

\begin{figure}[htbp]
    \centering
    \includegraphics[scale=0.28]{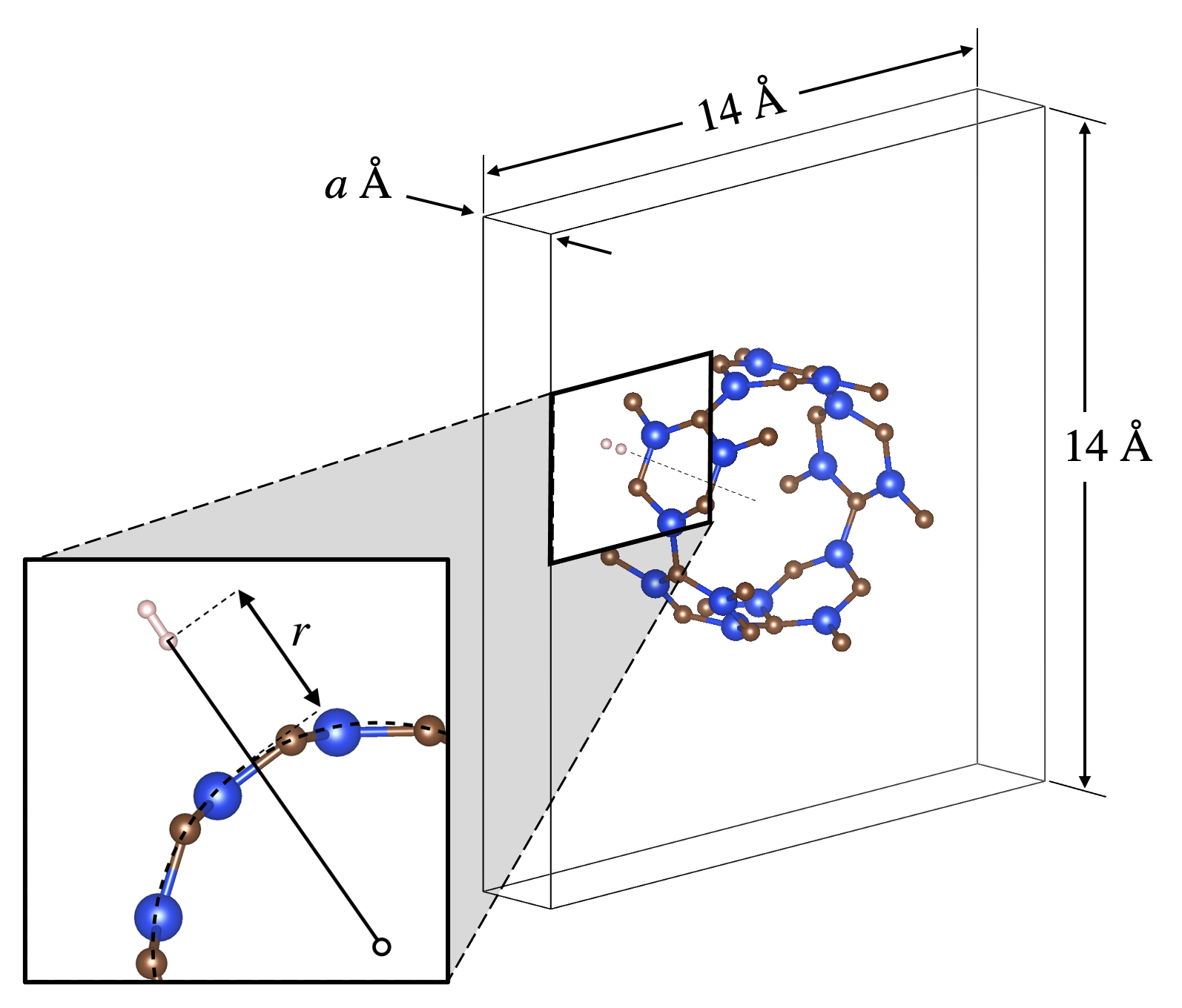}
    \caption{\label{fig:geom}
      A vertically oriented hydrogen molecule over an SiCNT hollow site.
      Inset: the distance is measured over the largest diameter of the SiCNT,
      which corresponds to the C site.
    }
\end{figure}

We selected the type-I (5,5) armchair SiCNT
with 20 atoms per primitive cell.
Consisting only of Si--C bonds, type-I 
was found to be more stable than those containing
Si--Si and C--C bonds~\cite{2007ALA, 2008ALA, 2004MEN}. 
The basic structure was built using
a generic nanotube generation tool~\cite{tubegen}
and later optimized in DFT.\@
A $14{\times}14{\times}a$ \AA{} simulation cell was used
to minimize spurious interactions with periodic images,
and the periodic unit length ($a$) was optimized along with the other geometries.
We placed a hydrogen molecule with a fixed vertical orientation
over an SiCNT hollow site (Figure~\ref{fig:geom}).
The distance was varied to obtain a binding curve,
for which the energies were least-square-fitted to a Morse potential.
\vspace{2mm}
The binding energy corresponding to the energy change from the free
to the bound configuration is defined as
\begin{equation}
  \label{eq:bc}
  \Delta E(r) = E_\mathrm{H_2-SiCNT}(r) - E_\mathrm{H_2} - E_\mathrm{SiCNT}.
\end{equation}
Each component on the right-hand side
was computed independently using the same simulation cell.
Since the diameter of the SiCNT is not singular
due to lengthwise buckling,
we defined $r$ as the distance from the hydrogen molecule to the center of the nanotube
subtracted by the largest diameter within the primitive cell.

\subsection{Density Functional Theory (DFT)}\label{dft}

The PWSCF binary in the QUANTUM ESPRESSO~\cite{qe} package
was used for all DFT calculations
based on plane wave basis sets with pseudopotentials.
The kinetic energy cutoff and k-point grid size were converged
at 150 hartree and a $1{\times}1{\times}6$
Monkhorst--Pack grid~\cite{1976MON}, respectively, 
to achieve chemical accuracy.
To reduce the time-step error in the later DMC stage,
the nonsingular energy-consistent pseudopotential reported by
Burkatzki, Filippi, and Dolg~\cite{2007BUR}
was utilized for all calculations.
All geometry optimizations were performed
within PBE with total force and energy thresholds
of $10^{-10}$ and $10^{-4}$ a.u., respectively.

\vspace{2mm}
We compared the vdW approaches DFT-D2, DFT-D3, vdW-DF2, and 
rVV10~\cite{2004DIO,2006GRI,2010GRI,2010LEE,2013SAB,2015THO},
of which the first two are based on an explicit $R^{-6}$ pairwise potential
operating on atomic coordinates,
whereas the latter have an exchange-correlation potential
modified by the addition of a nonlocal correlation term.
The additional term in the exchange-correlation potential
allows for changes in the charge density,
providing more information for analyzing the binding formation.
The local density functional PZ~\cite{pz}
and gradient-corrected PBE~\cite{pbe}
were employed as baselines.

\subsection{Diffusion Monte Carlo (DMC)}

\vspace{2mm}
QMCPACK~\cite{qmcpack} was used for all quantum Monte Carlo (QMC) calculations,
the main DMC calculations, and for the variational Monte Carlo (VMC) calculations 
used in preparation of the trial wave function.
The trial wave function was a Slater--Jastrow type comprising a single determinant 
with PBE-DFT orbitals and a Jastrow factor with optimized parameters at the VMC level.
Parameters consisting of one- and two-body interaction terms were used,
each in the form of B-splines with 10 optimizable parameters for each atom type.
For efficiency, Jastrow parameters were optimized using a hybrid method
mixing the linear method algorithm~\cite{Toulouse2007}
and a accelerated descent method as described by Otis and Neuscamann~\cite{2019OTI};
The orbitals were pre-projected into B-splines~\cite{2004ALF}
to increase the computational efficiency.
The pseudopotential parts were evaluated with the $T$-move scheme~\cite{2010CAS}.
The time-step and finite size errors were eliminated through linear extrapolation,
first by the time-step and then by the real space supercell size.
The DMC time-steps were 0.0025, 0.01, and 0.04 a.u,
and the supercells were 2, 6, and 8 times duplicates of the primitive unit cell
in the cylinder length,
each with twist-averaged boundary conditions~\cite{2001LIN}
on regular grids of $1{\times}1{\times}8$, $1{\times}1{\times}3$, and $1{\times}1{\times}2$,
respectively.
The mixed boundary condition was applied with open boundaries on the 14 \AA{} sides,
and the target walker population was set to 4,096.

\section{Results and Discussion}\label{results_discussion}

\subsection{Structural properties}\label{structure}

We found that the geometry of the (5,5) SiCNT is relatively unaffected
by the choice of XC functional.
(Table~\ref{tab:opt_geom}).
This result is unsurprising due to its construction of
only single covalent bonds involving \textit{s} and \textit{p} electrons.
The obtained bond lengths were also consistent with those reported in prior
hybrid works~\cite{2008ALA,2017WAN},
albeit slightly shorter than those obtained utilizing cluster
models~\cite{2018TAB,2004MEN}.
It is not unusual to have a slight distortion near cluster terminations that
does not exist in periodic models such as the one used in this work.
Thus, we conclude that the conventional XCs are sufficient to describe the SiCNT
model used in this work.
\begin{table}[htbp]
  \centering
  \caption{
    LDA- and GGA-optimized structural parameters of a (5,5) SiCNT given in \AA{}\@.
    The subscript $d$ denotes diagonal bonds relative to the cylinder axis.
  }
  \vspace{2mm}
  \begin{tabular}{@{}llllll@{}}
    \toprule
    XC & $R_\mathrm{{SiCNT}}$ & Buckling &
    $\langle\textrm{Si--C}\rangle$ &
    $\langle\textrm{Si--C}_\textrm{p}\rangle$ &
    $\langle\textrm{Si--C}_\textrm{d}\rangle$  \\
    \midrule
    LDA & 4.341 & 0.103 & 1.791 & 1.790 & 1.793 \\
    GGA & 4.338 & 0.101 & 1.790 & 1.788 & 1.792 \\
    \bottomrule
  \end{tabular}
\label{tab:opt_geom}
\end{table}

\subsection{H$_2$ adsorption on SiCNT}

\begin{figure}[htbp]
  \centering
        \includegraphics[scale=0.55]{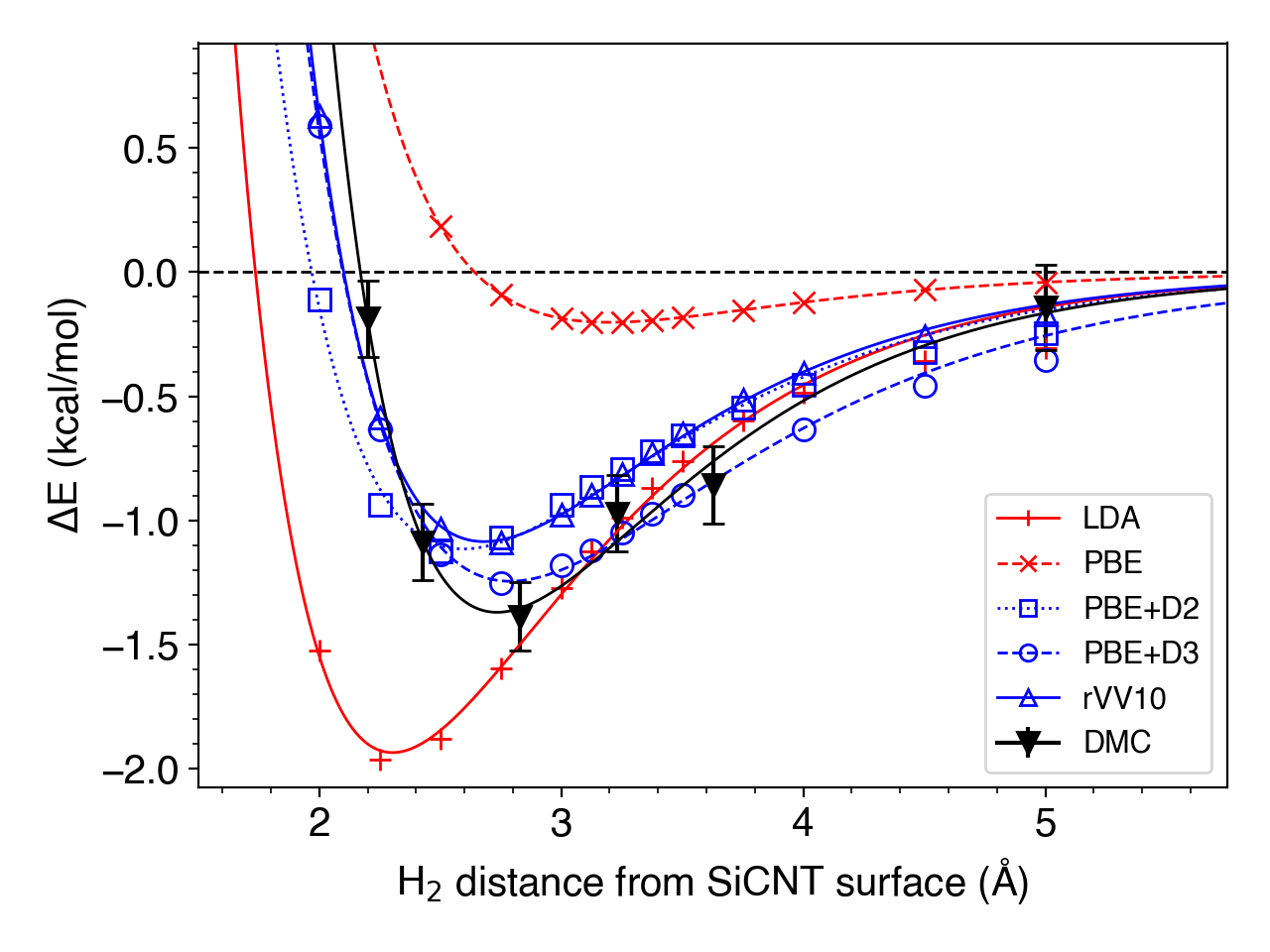}
    \caption{\label{fig:bc}
        Binding energy curves of H$_2$ on an SiCNT
        with evaluated XCs and DMC\@.
        Colored dashes indicate each curve's minimum.
        While energetically similar,
        the minima from the vdW-corrected XCs are at different locations,
        and they slightly underbind relative to DMC\@.
    }
\end{figure}
The main interest is how well the adsorption curves
from the vdW corrections agree with the DMC results
and among themselves.
As shown in Figure~\ref{fig:bc}, it is clear that LDA (GGA)
severely over (under)-binds the DMC target value by
25 (51) meV, or 0.6 (1.2) kcal/mol.
This result is consistent with the established behavior of LDA and GGA
in vdW-dominated systems.
The non-empirical vdW-DF2 and rVV10 functionals return virtually identical binding energies,
but their H$_2$ separations differ by approximately 0.2 \AA{}.
Interestingly, the D2 energetics agree more with the non-empirical methods
relative to D3.
This was accompanied by a slight underestimation of the H$_2$ separation distance,
which was corrected by D3.
Considering they both start from PBE energies, this is satisfactory.
In general, all vdW corrections underbind relative to DMC
but with more reasonable adsorption minima.
This suggests that while the geometries derived from these functionals can generally be trusted,
a more careful consideration of their energetics is necessary.

\begin{figure}[htbp]
  \begin{center}
     \subfigure[LDA]{\fbox{\includegraphics[scale=0.18,clip]{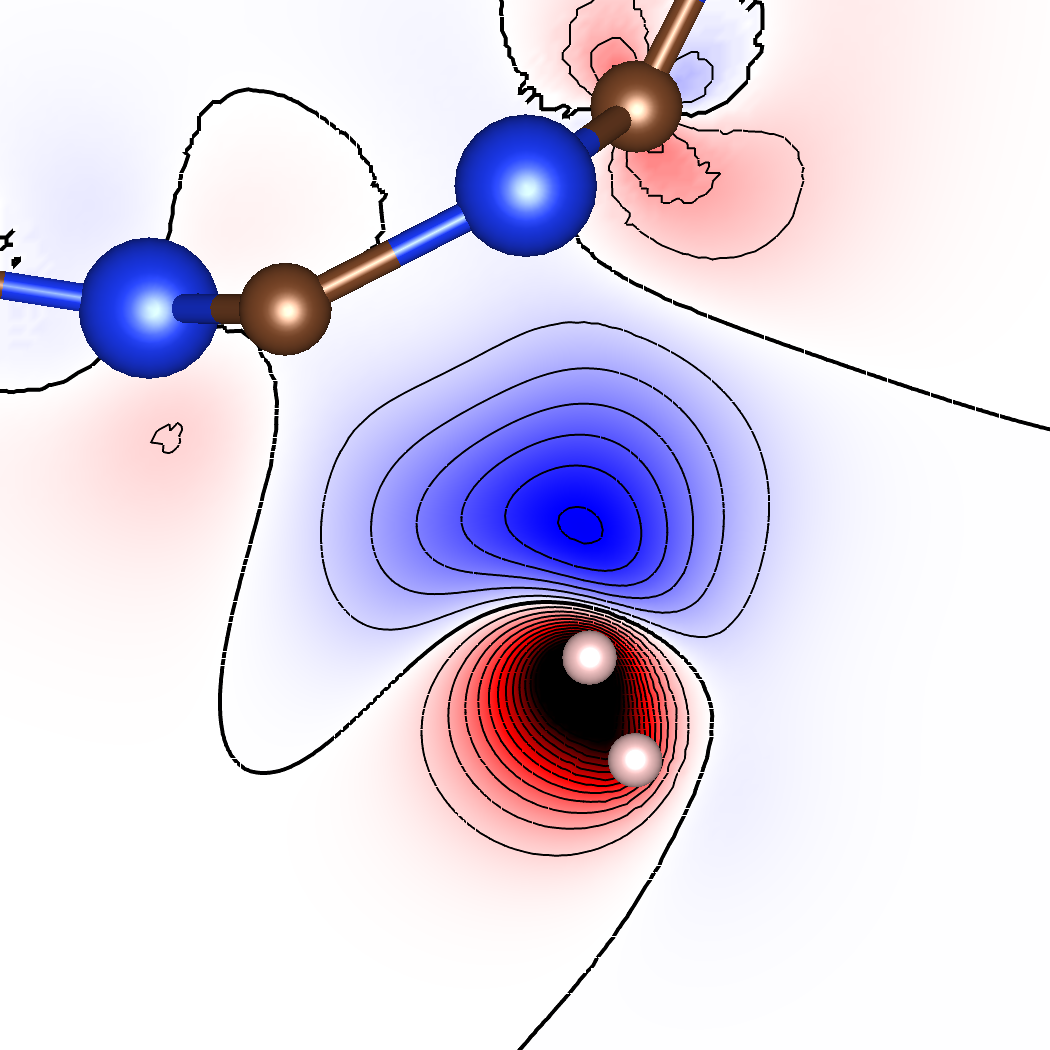}}}
     \subfigure[PBE]{\fbox{\includegraphics[scale=0.18,clip]{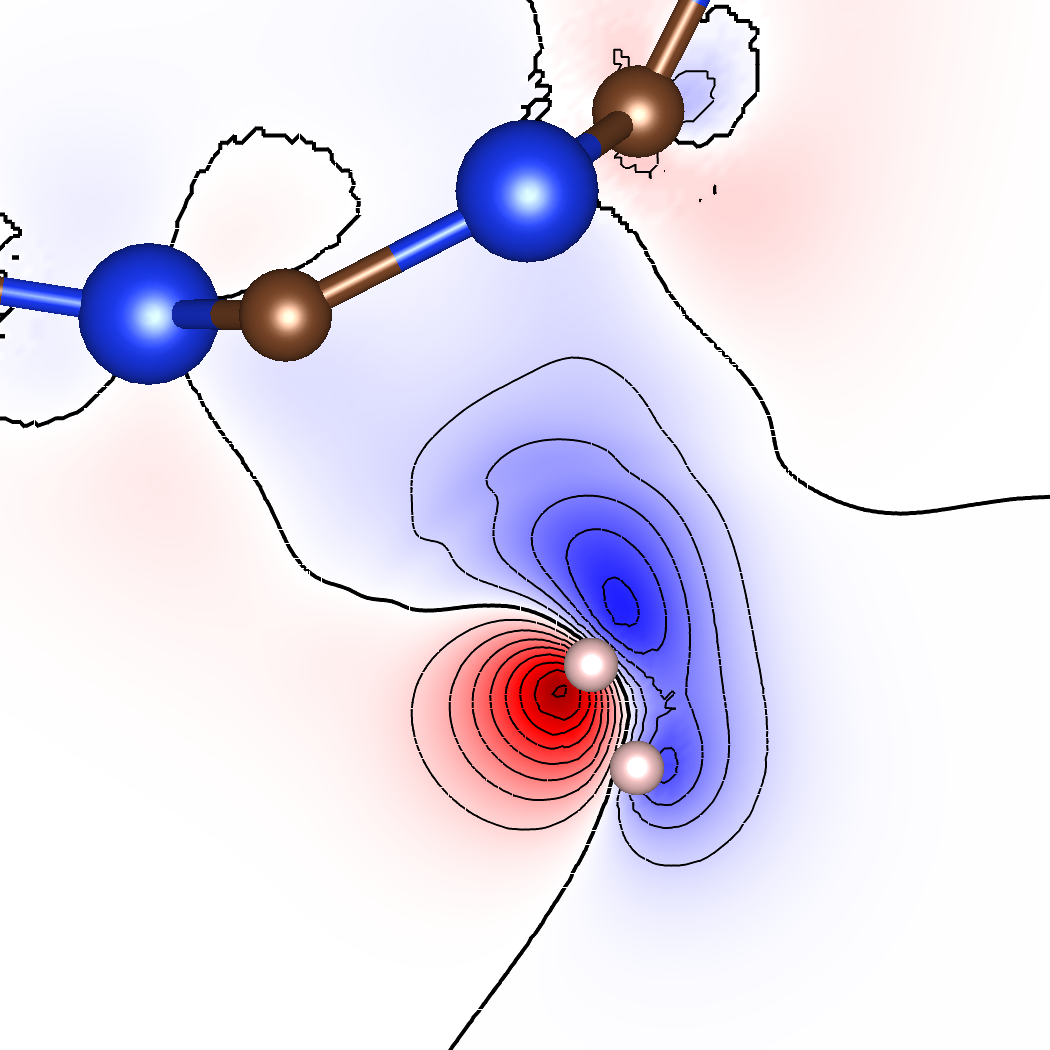}}}
     \subfigure[vdW-DF2]{\fbox{\includegraphics[scale=0.18,clip]{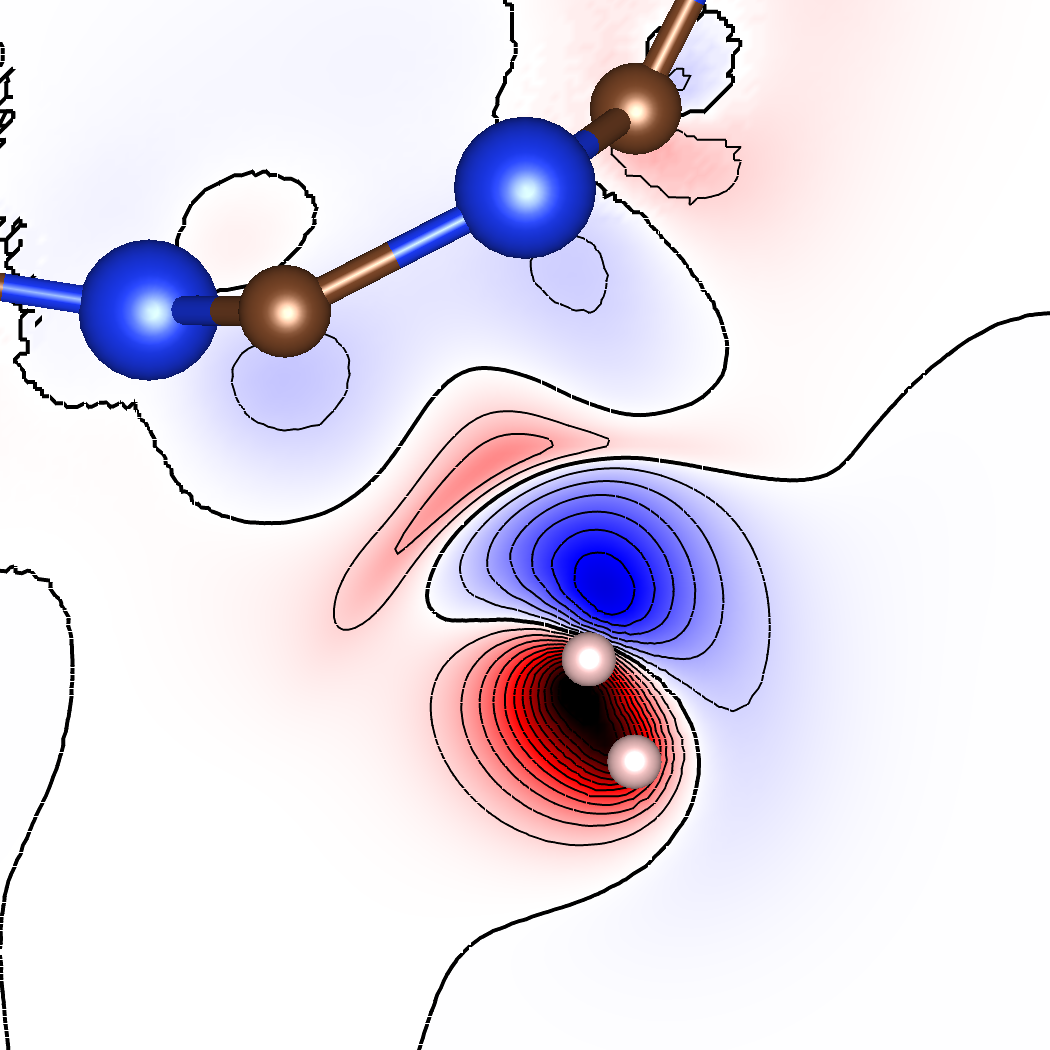}}}
     \subfigure[rVV10]{\fbox{\includegraphics[scale=0.18,clip]{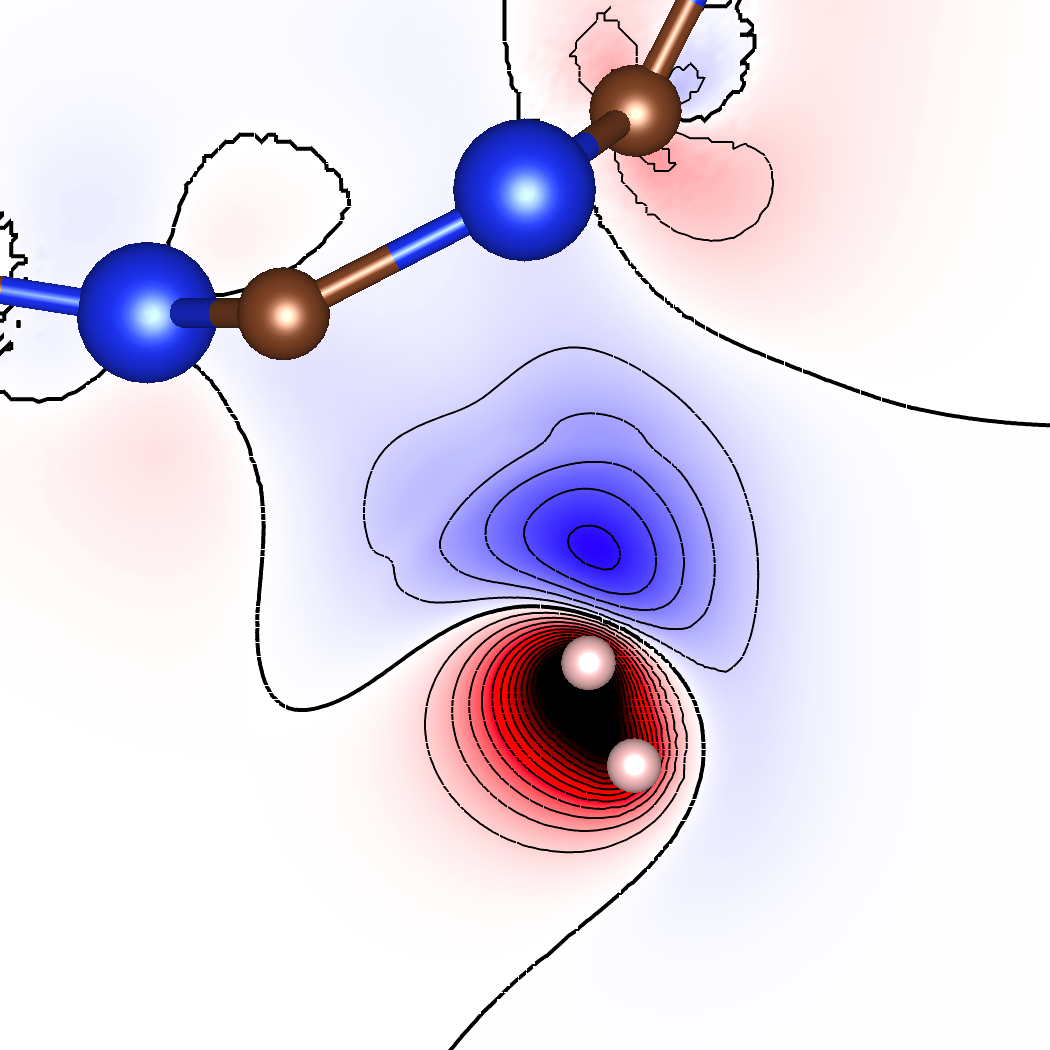}}}
  \end{center}
  \caption{\label{fig:cg}
    Charge density difference $\Delta\rho(r)$ in the presence
    of other fragment vs.\ isolated fragments:
    (a) LDA, (b) PBE, (c) vdW-DF2, and (d) rVV10.
    See the text for the definition of $\Delta\rho(r)$.
    Blue denotes a positive difference,
    while red denotes a negative difference.
    The distributions from vdW-DF2 and rVV10 are notably
    distinct; rVV10 is similar to LDA,
    while vdW-DF2 shows a slight dip in between
    despite their similar binding curves.
}
\end{figure}

\vspace{2mm}
LDA is known to produce spurious covalent bonding
between noncovalent molecules due to the self-interaction error~\cite{2010HON,2013HON}.
This gives rise to ca. 0.6 kcal/mol overbinding.
GGA improves the self-interaction error, but at the cost of weak intermolecular
interactions as the vdW interaction
is not inherently accounted for.
To illustrate this point,
we plotted the charge density difference between the whole and isolated components,
$\Delta\rho(r) = \rho_\mathrm{SiCNT+H_2}(r) - \rho_\mathrm{SiCNT}(r) - \rho_\mathrm{H_2}(r)$,
in Figure~\ref{fig:cg}.
The charge accumulation between the H$_2$ molecule and SiCNT,
i.e.,\ spurious covalent bonding,
is prominent for LDA as shown in Fig.~\ref{fig:cg} (a).
This feature is not present in the PBE data, as shown in Fig.~\ref{fig:cg} (b),
and is instead replaced by a much weaker
redistribution of charge toward the SiCNT surface.
Note that while the D2 and D3 corrections properly address the binding curve,
they do not affect the charge density at a given geometry.
In contrast, vdW-DF2 and rVV10 achieve their corrections
by deforming the charge density
(as a side effect of changes at the wavefunction level)
using non-local perturbations in the correlation integral.
As shown in Fig.~\ref{fig:cg} (c) and (d) for vdW-DF2 and rVV10, respectively,
there is a slight dip in charge density in the shape of a bow;
this is associated with the existence of noncovalent-type interactions~\cite{2011GAR},
although it is much weaker in the rVV10 case.
Interestingly, the rVV10 distribution is much closer to the LDA distribution
as opposed to the distinct distribution of vdW-DF2\@.
\vspace{2mm}
The lack of proper vdW interactions in LDA and GGA results in misestimation
by more than 0.5 kcal/mol,
with further potential for errors arising from incorrect geometry predictions.
Since the target adsorption energy is
3.5--11.5 kcal/mol,
this misestimation is not negligible.
We thus claim that the inclusion of vdW corrections in 
the XC functional is essential for the quantitative evaluation of adsorption energies,
particularly in the vdW-heavy H$_2$-SiCNT system,
and more generally for the further computational exploration of storage materials.

\subsection{Fully optimized geometries}

In Section 3.2, we varied only the distance between H$_2$ and SiCNT
for our benchmark purpose of evaluating the vdW interactions.
Here, we consider the fully optimized geometries
to evaluate a more realistic adsorption energy. 
Table~\ref{tab:adsorption_ge} gives
the structural information obtained
from the selected DFT methods.
Similar to the findings discussed in Section 3.2,
the vdW-corrected functionals clearly give highly similar trends,
while LDA (GGA) over (under)-binds.
Regarding the vdW-corrected functionals,
vdW-DF2 and rVV10 agree well with each other in terms of
the structural parameters and adsorption energy.
In particular, the surface angle determined by PBE+D3 
is markedly different from those determined by vdW-DF2 and rVV10. 
\begin{table}[h]
  \centering
  \caption{\label{tab:adsorption_ge}
  Optimized H$_2$ conformations and adsorption energies from all tested XC functionals.
  The energy values are from full optimization calculations.}
  \vspace{2mm}
  \begin{tabular}{@{}llll@{}}
    \toprule
    XC  & $R\mathrm{_{H_2}}$ [\AA] & $\theta_{s}$ [deg.]
    & $E_{\mathrm{ads}}$ [kcal/mol]\\
    \midrule
      LDA    & 2.573 & 51.3 & 2.375 \\
      GGA    & 3.437 & 37.6 & 0.291 \\
      PBE+D3 & 3.000 & 45.7 & 1.421 \\
      vdW-DF2& 3.170 & 54.4 & 1.268 \\
      rVV10  & 2.930 & 53.9 & 1.298 \\
    \bottomrule
  \end{tabular}
\end{table}
\vspace{2mm}
The above findings can also be investigated by the charge density
analysis shown in Figure~\ref{fig:cg}. 
Considering the change in the charge density distribution,
LDA and the two vdW-DFT functionals (vdW-DF2 and rVV10) give
different shapes than PBE (PBE+D3);
the surface angles $\theta_s$ for LDA, vdW-DF2, and rVV10 are
greater than 50$^{\circ}$, whereas PBE and PBE+D3 have smaller $\theta_s$ values.
Only the pairwise vdW corrections cannot reproduce
the structure properly, 
but the self-consistent charge deformation is also important
in obtaining proper structures.
As the charge density at the bonding region between
H$_2$ and SiCNT increases (LDA $>$ rVV10 $>$ vdW-DF2), 
$R_\mathrm{H_2}$ and $\theta_s$ both decrease
(LDA $<$ rVV10 $<$ vdW-DF2).
As mentioned in Section 3.2, 
the LDA bonding is not noncovalent but spurious.
It is evident from 
the correlation between the charge density and structure
that the LDA overbinding correlates to the highest
charge density at the bonding region.

\section{Conclusion}

We performed DMC calculations and various
DFT simulations
to evaluate the adsorption energies of a hydrogen molecule
on an SiCNT with a (5,5) armchair structure.
Recently developed XC functionals
designed to reproduce vdW interactions
(PBE+D3, vdW-DF2, and rVV10)
and conventional XC functionals (LDA-PZ, GGA-PBE, and B3LYP) 
were compared to DMC as a reference.
Overall, all the vdW-corrected XC functionals 
agree well with DMC, whereas LDA (GGA) over (under)-binds.
The self-consistent nonlocal 
correlation functionals, vdW-DF2 and rVV10, 
give almost the same adsorption energies.
Differences in the structural properties were found to closely
correlate with differences in the charge density distribution.
A higher charge density in the bonding region 
leads to a shorter distance between H$_2$ and SiCNT
and larger surface angle.
The magnitude of the vdW interaction was estimated
to be ca. 51 meV (1.2 kcal/mol), which corresponds to 9--29 \% of
the ideal adsorption energy for hydrogen storage.
This finding implies the importance of vdW corrections
within the framework of DFT.\@
We thus conclude that
protocols based on vdW-corrected XC functionals
will advance the computational investigation and exploration 
of storage materials in the near future.

\section*{Acknowledgments}

The computations in the present study were partially performed
using the facilities of the Research Center for Advanced Computing Infrastructure 
(RCACI) at JAIST and resources of
the Argonne Leadership Computing Facility,
which is a DOE Office of Science User Facility
supported under Contract No. DE-AC02-06CH11357 through an award provided 
by the Innovative and Novel Computational Impact
on Theory and Experiment (INCITE) program.
A.B. and H.S were supported by the U.S. Department of Energy,
Office of Science, Basic Energy Sciences,
Materials Sciences and Engineering Division, 
as part of the Computational Materials Sciences Program
and Center for Predictive Simulation of Functional Materials.
R.M. is grateful for financial supports from 
MEXT-KAKENHI (JP19H04692 and JP16KK0097), 
FLAGSHIP2020 (project nos.\ hp190169 and hp190167 at K-computer), 
the Air Force Office of Scientific Research 
(AFOSR-AOARD/FA2386-17-1-4049;FA2386-19-1-4015), 
and JSPS Bilateral Joint Projects (with India DST). 
K.H. is grateful for financial support from 
the HPCI System Research Project (Project ID:\ hp190169) and 
MEXT-KAKENHI (JP16H06439, JP17K17762, JP19K05029, and JP19H05169).
\bibliography{references}
\end{document}